\documentclass[sigconf]{acmart}

\usepackage{booktabs} 
\usepackage{algorithm}
\usepackage[noend]{algpseudocode}
\usepackage{caption}
\usepackage[margin=20pt]{subcaption}
\usepackage{hyperref}
\usepackage{enumerate}
\usepackage{mymacros}
\usepackage{balance} 
\usepackage{paralist}
\usepackage{dsfont}
\usepackage{rotating}
\usepackage{multirow}
\usepackage{dblfloatfix}

\usepackage[font={small},skip=0.5pt]{caption}
\usepackage{setspace}

\usepackage{graphicx}
\usepackage[normalem]{ulem}
\useunder{\uline}{\ul}{}

\usepackage{amsthm}
\theoremstyle{definition}
\newtheorem{remark}{Remark}

\copyrightyear{2021}
\acmYear{2021}
\setcopyright{acmcopyright}\acmConference[KDD '21]{Proceedings of the 27th ACM
SIGKDD Conference on Knowledge Discovery and Data Mining}{August 14--18,
2021}{Virtual Event, Singapore}
\acmBooktitle{Proceedings of the 27th ACM SIGKDD Conference on Knowledge
Discovery and Data Mining (KDD '21), August 14--18, 2021, Virtual Event, Singapore}
\acmPrice{15.00}
\acmDOI{10.1145/3447548.3467129}
\acmISBN{978-1-4503-8332-5/21/08}

\renewcommand\footnotetextcopyrightpermission[1]{} 

\begin{document}
\fancyhead{}
\title{On Post-selection Inference in A/B Testing}


    
\author{Alex Deng}

\authornote{Work primarily done when working at Microsoft}
\authornote{First two authors contributed equally to this work. }
\affiliation{
  \institution{Airbnb}
  \city{San Francisco} 
  \state{CA} 
    \country{USA}
  \postcode{98052}
}
\email{alex.deng@airbnb.com}

\author{Yicheng Li}
\authornotemark[2]
\affiliation{
  \institution{Microsoft Corporation}
  \city{Redmond} 
  \state{WA} 
    \country{USA}
  \postcode{98052}
}
\email{yicl@microsoft.com}

\author{Jiannan Lu}
\affiliation{
  \institution{Microsoft Corporation}
  \city{Redmond} 
  \state{WA} 
    \country{USA}
  \postcode{98052}
}
\email{jiannl@microsoft.com}

\author{Vivek Ramamurthy}
\affiliation{
  \institution{Microsoft Corporation}
  \city{Redmond} 
  \state{WA} 
  \country{USA}
  \postcode{98052}
}
\email{viramam@microsoft.com}

\begin{abstract}
When interpreting A/B tests, we typically focus only on the statistically significant results and take them by face value. This practice, termed post-selection inference in the statistical literature, may negatively affect both point estimation and uncertainty quantification, and therefore hinder trustworthy decision making in A/B testing. To address this issue, in this paper we explore two seemingly unrelated paths, one based on supervised machine learning and the other on empirical Bayes, and propose post-selection inferential approaches that combine the strengths of both. Through large-scale simulated and empirical examples, we demonstrate that our proposed methodologies stand out among other existing ones in both reducing post-selection biases and improving confidence interval coverage rates, and discuss how they can be conveniently adjusted to real-life scenarios. 
\end{abstract}


\settopmatter{printfolios=true}
\maketitle

\section{Introduction}
\label{sec:intro}

\subsection{Background}

Statistical inference, a major force behind the big data revolution, builds bridges between massive amounts of data and the probabilistic models governing their underlying generating processes, and enables transformation of learning from one dataset to more general populations \citep{casella2002statistical,murphy2012machine}. Typically, the goal of statistical inference is to infer quantities associated with the probabilistic models, e.g., common descriptive statistics (mean, median etc.), trained/fitted machine learning model parameters, or model evaluation metrics (accuracy, error rate etc.). In particular, the main output of the inference consists of a \emph{point estimation} and its corresponding \emph{confidence interval} \citep{lehmann2006theory}, representing both a prediction of the unknown quantity's value and the associated uncertainty. Due to the well-known duality, inferences yielding confidence intervals with nominal coverage rates implies hypothesis tests with proper type-I error rates \citep{lehmann2006testing}. 

On-line controlled experiments (a.k.a A/B tests) are widely used to evaluate and optimize web products, e.g., search engines \citep{kohavi2009controlled,googlesurvey,budylin2018consistent}, social networks \citep{xu2015infrastructure}, web streaming services \citep{xie2016improving,turnbull2019learning} and shared economy platforms \citep{garcia2018understanding}. At its core, A/B testing aims at inferring the treatment effects of new experiences and features on a set of metrics. Typically, collecting feedback from users interacting with web products \citep{abScale} is cost-efficient and near real-time, opening up both the potential opportunities and the challenges of large-scale A/B tests. First, the amount of data for each experiment can be large. This is challenging for computation but a blessing for analysis, as it allows the Central Limit Theorem \citep{asympstat} to kick in, and we don't need to make strong assumptions on the data generating process \citep{Dengkdd2018}. Second, the number of analyses required for each experiment can be large. Experimenters are often interested in a set of metrics, ranging from tens to thousands \citep{Deng:2016b}. Moreover, each metric can be analyzed for different segments such as markets, operating systems and so on \cite{Gupta:2019,deng2016concise}. Third, the number of experiments conducted during a release cycle is large. The types of changes teams make in a feature/treatment in those iterations range from minor configuration changes to complete rewrites.

\subsection{Post-selection Inference in A/B testing}
\label{sec:intropostinf}

Post-selection inference naturally arises in simultaneous analyses. We consider a standard A/B test with a treatment and a control group of sample sizes $N_T$ and $N_C$ and metric values $Y_T$ and $Y_C$, respectively. A metric could be in the form of an average across i.i.d. samples, but is not limited to it. The central limit theorem entails that when sample sizes are large enough, the estimated treatment effect 
$
\Delta = Y_T-Y_C
$ 
approximately follows a Gaussian distribution with mean $\mu$ and variance 
$
\sigma_T^2/N_T+\sigma_C^2/N_C.
$
With i.i.d. observations, $\sigma_T^2$ and $\sigma_C^2$ are sample variances of the respective groups. With non i.i.d. observations, we need to leverage more advanced methods (e.g., the Delta method \citep{Deng:2017}) to properly calculate the sample variances. We define the effective sample size and pooled variance as 
\begin{equation*}
N = \left (\frac{1}{N_T}+\frac{1}{N_C}\right)^{-1},
\quad
\sigma^2 = N \left( \frac{\sigma_T^2}{N_T} + \frac{\sigma^2_C}{N_C} \right).
\end{equation*}
Consequently,
$
\Delta \sim \text{Normal}(\mu,\sigma^2/N).
$
Our goal is to infer the average treatment effect $\mu$, and $\Delta$ is an unbiased estimator of $\mu$. However, if we repeatedly sample many times, and only report if $\Delta>\mu+1.65\sigma/\sqrt{N}$,\footnote{$\mu$ is unknown in reality. This example is just for illustration. } then by the tail probability, 5\% of the cases would meet this criterion, all of which would over-estimate $\mu$.


This phenomenon is ubiquitous in modern data analysis. Indeed, most statistical theories require us to pre-specify a scientific question and then provide an answer, whether it's ``favorable'' or not. However, in the post-selection scenario we ask multiple questions, and choose to answer a subset after peeking at the data. In A/B testing, in the presence of many metrics, segments, or treatments being tested, even experts often filter down to only statistically significant results, thus introducing biases. Another typical selection is to continuously monitor results and stop collecting data when results are favorable \citep{Deng:2016a,ju2019sequential}. Such practices seem to be sound procedures, lest we be drowned by an ocean of noisy numbers. At the same time, they are the epitome of post-selection inference. Intuitively, they introduce bias  (also known as ``winner's curse'', see \citep{lee2018winner}), because we tend to select questions for which the data provides ``favorable'' answers. Unfortunately, as intuitive as it sounds, assessing the post-selection bias seems impossible for real-life data-sets, because we don't know the ground-truth we seek to estimate. 

\begin{figure}[!htbp]
    \centering
    \includegraphics[width=\columnwidth]{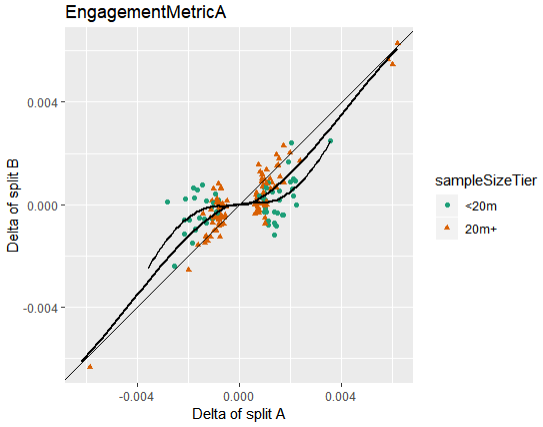}
    \caption{Each point represents a pair of scaled observed effects ($\Delta$s) of a user engagement metric from a random split of the same experiment, and can be seen as two independent replications of the same experiment. $p-$value selection (< $0.1$) is applied to split A and 168 out of 1026 experiments are selected and shown. 
    Split B's Delta is unbiased without selection, regression of split B on split A gives a sense of how split A's Delta is biased. Two smoothed regression lines for small and large sample sizes both show strong \emph{non-linearity} of the true regression and larger selection bias when sample sizes are small.}
    \label{fig:illust}
\end{figure}

An attainable alternative to assess post-selection bias is through replication. For a given post-selected estimate, if we re-run the exact same experiment and conduct the same analysis without selection for the replication, the new estimate should be unbiased. By comparing the two estimates, we can assess the bias. In practice, a pseudo-replication pair can be formed by randomly splitting the experiment traffic into two splits $A$ and $B$ \citep{coey2019improving}, where split $A$ is treated as the first run, and split $B$ is its replication. For each experiment, we evaluate the scaled observed effect (i.e., $\Delta$ standardized by its standard deviation) for both splits $A$ and $B$. Figure~\ref{fig:illust} shows the scaled observed effect pairs $(\Delta_A, \Delta_B)$ of 168 experiments, out of 1026 experiments conducted by the same product team selected by the criterion of $p-$value<0.1. We group the selected 168 experiments by whether their sample sizes exceed twenty million\footnote{We explored different grouping mechanisms, all of which yielded similar results.}, on which we fit two separate smooth local regression curves. If both $\Delta_A$ and $\Delta_B$ are unbiased with independent noises, the observations should be along the reference line $\Delta_B=\Delta_A$. However, both curves are under/above the reference line for positive/negative values of $\Delta_A$. We know $\Delta_B$ is unbiased because we didn't select based on it. Then $\Delta_A$ must be upwardly biased when positive and downwardly biased when negative. Moreover, the adjustment as illustrated by the local regression curve is \emph{non-linear} --- both curves appear flat near 0 and seem to approach $\Delta_B = \Delta_A$ asymptotically. In other words, the larger $\Delta_A$ is, the lesser is the adjustment needed. In addition, experiment sample size also affects the adjustment, as seen from the two fitted smoothing curves. This hinders the direct curve fitting since we don't want the estimates to depend on the subjective choice of grouping (20 million samples in this example). We will discuss more about experiment splitting and regression of these pairs in Section \ref{sec:es} as we survey related works.

The key insights from Figure~\ref{fig:illust} are:
\begin{compactenum}
    \item selection based on noisy observations leads to biased estimate;
    \item the amount of bias depends on the signal-noise-ratio and could be non-linear and sample-size-dependent.
\end{compactenum}

In this paper, we do emphasize identifying the pitfalls of post-selection and assessing its induced bias. The new methodologies we propose, are aimed at rendering trustworthy treatment effect estimation in A/B tests, that is immune to the post-selection bias.

\subsection{Contributions and Organization}
\label{sec:contribution}
Beyond raising awareness of the necessity for trustworthy post-selection inference, this paper makes the following contributions. First, we comprehensively survey the existing literature and provide a holistic view that facilitates development of new methods. Second, we propose two new methods, both of which significantly improve performance. Third, we present a solution for the cold-start scenario, when only a small amount of historical experiment data exists. Fourth, we conduct extensive simulation and empirical studies to demonstrate the advantages of our proposed methodologies. To the best of our knowledge, we are the first to evaluate both post-selection biases and confidence interval coverage rates, using real-life experiments. The new empirical Bayes method we propose, is shown to provide, both accurate point-estimation and confidence intervals. It is also adaptive to different treatment effect distributions. Finally, we are sharing our implementations with the community for replication and adoption. 

This paper is organized as follows. Section~\ref{sec:notation} reviews existing post-selection methods. Sections~\ref{sec:newhope} and \ref{sec:extra} introduce two new methods for the post-selection adjustment, and propose a Bayes Factor bound method for cold-start scenarios. Sections \ref{sec:simulation} and \ref{sec:empstudy} highlight the advantages of our proposed methods via simulated and empirical examples, respectively. Section~\ref{sec:conclusion} concludes and discusses future work. We provide R implementations of new methods and simulation procedures, for reproduction\footnote{https://aka.ms/exp/code/abpostselection}.

\section{Survey of Existing work}
\label{sec:notation}

\subsection{Conditional Maximum Likelihood}
\label{sec:cmle}
A conditional maximum likelihood estimate (CMLE) is a traditional maximum likelihood inference conducted under some selection criterion. More specifically, the distribution under the null hypothesis is conditioned on the action of selection and the inference is conducted accordingly. \citet{reid2017post} studied the inference of a Gaussian mean $\mu$ with known variance $\sigma^2,$ under the selection criteria $|\Delta| \ge K,$ a setup close to A/B testing. In this case, the CMLE is a solution of $\mu$ to the following equation 
\begin{equation}\label{eq:cmle}
    \Delta - \mu = \sigma \frac{\phi\left(\frac{K-\mu}{\sigma}\right) - \phi\left (\frac{-K-\mu}{\sigma}\right)}{\Phi\left(\frac{-K-\mu}{\sigma}\right)+1-\Phi\left(\frac{K-\mu}{\sigma}\right)}\ .
\end{equation}
This CMLE has an intuitive explanation of iterative bias correction. If we know $\mu$, then the expected selection bias $\e (\Delta - \mu | |\Delta|>K)$ is the right hand side of \eqref{eq:cmle}. Because we don't know $\mu$, we equate the expected bias to the observed bias $\Delta - \mu$ and obtain \eqref{eq:cmle}, by means of an iterative procedure. 

\begin{remark}
\citet{lee2018winner} applied expected bias correction to A/B testing, focusing on the marginal expectation 
$
\e [(\Delta - \mu) \mathds{1}_{|\Delta|>K}]
$ 
instead of the conditional expectation, and stopping at the first iteration. Consequently, their correction can be applied to a compound estimation of a group of estimates, not to each individual. 
\end{remark}

\subsection{Experiment Splitting}
\label{sec:es}

\citet{coey2019improving} pointed out that we could leverage the data points in Figure~\ref{fig:illust} (including those not passing the selection criterion) to train a regression model to predict $\Delta_B$ by $\Delta_A$. Because
$$
\e (\Delta_B|\Delta_A) 
= 
\e (\mu|\Delta_A),
$$
we immediately obtain a predictive model of $\mu$ given $\Delta_A.$ Because the regression is conditioned on observations, it takes post-selection into account, similar to CMLE. Consequently, data splitting naturally transforms post-selection inference into a standard supervised learning problem. However, a notable missing piece is the proper functional form of the learner that can capture the non-linear pattern we saw in Figure~\ref{fig:illust}. Linear or not, the predictive model should depend on the sample size of each experiment (\citet{coey2019improving} trained on the split data when all experiments have similar sample sizes). The trained model then needs to be properly ``scaled up to full sample size'' when making predictions for future experiments.

\subsection{Empirical Bayes}
\label{sec:eb}

Bayesian methods are known to be immune to post-selection bias, by conditioning on the observations \citep{efron2010large,efron2011tweedie,senn2008note,Lu:2016,Deng:2016a}. Consider a prior distribution
$
\mu\sim \pi
$
and the subsequent data-generating process 
$
\Delta \mid \mu \sim \text{Normal}(\mu, \sigma^2/N).
$ 
We can compute the posterior mean
$
\e (\mu|\Delta)
$ 
by the classic Tweedie's formula \citep{efron2011tweedie}. To be specific, let $l$ be the marginal log-likelihood of $\Delta$ (with fixed sample size $N$),
\begin{equation}
\label{eq:tweedie}
\e (\mu|\Delta) 
= 
\Delta + \frac{\sigma^2}{N} l^\prime (\Delta),
\quad
\var (\mu|\Delta) 
= 
\frac{\sigma^2}{N} 
\left\{
1+\frac{\sigma^2}{N} l''(\Delta)
\right\}.
\end{equation}
In particular, if 
$
\mu \sim \textrm{Normal}(0, \tau^2),
$
then
$
l'(\Delta)
$ 
is a linear function of $\Delta$ with the posterior mean 
\begin{equation}
\label{eq:js}
\e (\mu|\Delta) 
= 
\frac{\tau^2}{\sigma^2/N+\tau^2}\Delta.
\end{equation}
In practice, we can use historical A/B tests to estimate $\tau$ \citep{Deng:2015,dimmery2019shrinkage} and obtain the James-Stein shrinkage estimator \citep{efronmorris1973}. Notice that, although the shrinkage from \eqref{eq:js} is linear in $\Delta$, the shrinkage factor depends on the sample size in a non-linear way, so this method gives a very different estimator from fitting linear regression on split data. But Gaussian prior is a strong assumption. A more palatable assumption for the prior is uni-modal with slowly decaying tails, always shrinking $\Delta$ toward zero. When the tail of the prior is heavier than Gaussian, the adjustment is smaller for larger effects. This proposal is further supported by Figure~\ref{fig:illust}, where the empirically fitted curves are flat near zero, with increasing slope and therefore less adjustments for bigger $\Delta$s. Similar findings suggesting effect distribution having heavier-than-Gaussian tail were published previously \citep{azevedo2019empirical,goldberg2017decision}. In other words, we almost don't need to discount break-through features with large treatment effects.

The salient challenge of Bayesian methods is the specification of prior. This problem is partially alleviated by using real-life A/B tests to search within a family of prior distributions (Empirical Bayes). But then the challenge is to find a family of priors that can cover a large space of possibilities. \citet{efron2011tweedie} suggested directly estimating $l'(\Delta)$ nonparametrically from observed $\Delta$s. Unfortunately, this idea does not apply when sample sizes vary and $\Delta$ given $\mu$ is heteroskedastic. NEST method proposed by \citep{fu2020nonparametric} addressed the heteroskedasticity issue by extending Tweedie's formula to a multivariate version and applying a two-dimensional Gaussian kernel that weights observations by the distances in both location and scale. This new model-free empirical Bayes method looks promising, but may require a large amount of training data due to its nonparametric nature. Another practical challenge is that experimenters sometimes question whether the historical experiments used to train the priors can properly represent a new feature they are currently A/B testing. 

\section{New methodologies}
\label{sec:newhope}

\subsection{Motivation}
We surveyed conditional maximum likelihood estimator (CMLE), regression with experiment splitting (RwES) and empirical Bayes (EB). Among them, CMLE appears rather limited, as it conditions on a prescribed selection criterion instead of the observations. More importantly, practitioners often adopt fluid selection criteria (e.g., gradually changing the $p-$value threshold) when analyzing A/B tests. Simulation studies in later sections also show the inferiority of CMLE. RwES and EB both condition directly on the observations and share the same end goal of directly modeling $\e (\mu|\Delta).$ On one hand, EB is a ``generative'' method, as it models the prior $\pi,$ which subsequently determines $\e (\mu|\Delta).$ On the other hand, RwES is a ``discriminative'' method, as it takes the shortcut and directly models $\e (\mu|\Delta)$ by regression. Both RwES using simple linear regression and EB with Tweedie's formula (\ref{eq:tweedie}) lack the ability to account for the non-linearity with respect to the experiment sample size $N$ and/or the treatment effect $\Delta$. In the following sections, we propose an RwES-based and an EB-based method that provide straight-forward solutions to address the issue above. Finally, we unify the two frameworks into one, to provide more flexibility and usability to practitioners from different fields.

\subsection{TARwES: Theory-Assisted Regression}\label{sec:tareg}
How can we assist RwES with a non-linear functional form? Empirical Bayes! The EB method can be treated as a feature generator to help formulate the non-linear functional form. We call this hybrid approach \emph{Theory-Assisted Regression with Experiment Splitting (TARwES)}. As the name suggests, we use posterior mean formulas like \eqref{eq:js} from Empirical Bayes with various prior distributions as predictive features in the RwES method. These features are combined with the unadjusted observation $\Delta$ in the RwES step to fit a linear regression. For better regularity, we symmetrize the training data in RwES by mirroring each training data point (negative to positive, and vice versa). The regression model also does not include an intercept, meaning when $\Delta$ is 0 the prediction will always be 0. We emphasize that regularization is an absolute \emph{must} in TARwES, because those EB based features are all trying to predict the same ground-truth effect $\mu$. Different prior assumptions render these predictions different; some may fit large effects better, while some may fit small effects better. Nevertheless, these feature will be highly correlated. In the replication code, we implemented ridge regression and non-negative least squares \citep{slawski2013non}. 

Potential prior candidates for the EB based features should have the following two characteristics. First, they should capture both thin-tailed and heavy-tailed scenarios. Second, to facilitate large-scale computation, their exact or approximated posterior means should exist in closed-forms. To fulfill these criteria, we used EB with Gaussian and Laplace prior. While the Gaussian prior satisfies the thin-tailed scenario, we choose the Laplace prior with mean 0 and variance $\nu^2$ to cover the heavy-tailed one, which is also an important component of another new method proposed in the next section. Not only does the Laplace prior yield a closed form posterior mean and variance under Gaussian noise, i.e.,
\begin{align}\label{eq:lapreg}
    \e (\mu|\Delta) &= w(\Delta)(\Delta+b) + (1-w(\Delta))(\Delta - b),\\
   \var(\mu|\Delta) &=  \frac{\sigma^2}{N} - \frac{4\sigma^4}{N^2\nu^2}\frac{(F(\Delta)+F(-\Delta))f(\Delta)-2F(\Delta)F(-\Delta)}{(F(\Delta)+F(-\Delta))^2},
\end{align}
where 
\begin{align}
    b &= \frac{\sigma^2\sqrt{2}}{N\nu} \ , \quad  w(\Delta) = F(\Delta)/(F(\Delta)+F(-\Delta)), \\
    F(\Delta) &= \exp \left(\frac{\sqrt{2}\Delta}{\nu}\right) \Phi\left( \frac{\sqrt{N}}{\sigma}(-\Delta-b) \right), \ \Phi - \text{Gaussian CDF},\\
     f(\Delta) &= \frac{\nu\sqrt{N}}{\sigma\sqrt{2}}\exp \left(\frac{\sqrt{2}\Delta}{\nu}\right) \phi\left( \frac{\sqrt{N}}{\sigma}(-\Delta-b) \right), \ \phi - \text{Gaussian PDF},\
\end{align}
it generates \emph{bounded bias correction}, i.e., for large positive $\Delta$, the regression prediction will move close to 
$
\Delta - \sqrt{2}\sigma^2 / (N\nu)
$ 
and for negative $\Delta$, the asymptote is 
$
\Delta + \sqrt{2}\sigma^2 / (N\nu)
$
, which fits well with the observation made from Figure \ref{fig:illust}. In contrast, the linear multiplier correction property from a Gaussian prior yields a correction that grows as $|\Delta|$ increases boundlessly. Moreover, we found that t-prior based features, the other potential heavy-tailed candidates, are less useful compared to Laplace features through simulated and empirical studies \footnote{Therefore, going forward we exclude t-priors in TARwES.}. We refer readers to \citep{pericchi1992exact} for details of posterior mean and variance inference of a Laplace and a t-prior with a Gaussian noise. 

As a two-step hybrid method, TARwES combines the strengths of RwES and EB. First, EB relies heavily on the choices of priors, but in TARwES they are merely used as feature generators. We can use multiple priors, each of which provides a possible non-linear form of the regression function. The regression step using experiment splitting can empirically pick the best combination. This makes it more robust against prior mis-specification. Second, RwES cannot capture the non-linearity of the adjustment w.r.t. sample size, but TARwES captures this non-linearity within all the EB features because EB predictions like \eqref{eq:js} already incorporate the effect of sample size. We no longer need data hungry non-linear regression methods like gradient boosted trees in the regression step. In fact, we rely on EB features to capture most of the non-linearity, so in the regression step we only need to perform simple linear regression (with regularization). Also, we can train on split data and apply the model on full data without the ``scale up'' problem RwES faces\citep{coey2019improving}. Third, some prior parameters in EB method can be very hard to estimate. For example, the degrees of freedom of a t distribution is hard to estimate when it is small and the tail is heavy (the effective samples are only those at the tails). Instead of estimating these parameters, TARwES method allows us to treat these unknown parameters simply as different features and lets the regression step to optimize them.  

\subsection{Ghidorah: The Three-Headed Monster}\label{sec:thm}
In this section, we explore the direction of improving existing EB methods to match the observations made in Figure \ref{fig:illust}. Instead of directly modeling $\e(\mu|\Delta)$ as in the TARwES method (introduced in the previous section), we model the prior $\pi$ first and infer the induced estimate for $\e(\mu|\Delta)$ by applying the multivariate Tweedie's formula (cf. \citep{fu2020nonparametric}). There are two noticeable advantages of EB methods over the RwES methods. First, EB does not require data splitting. Also, it generates more numerically stable variance prediction, which will be detailed in Section \ref{sec:ci}. 

Motivated by the "prior fusing" idea, we propose a special prior. As the name indicates, the \emph{Ghidorah}\footnote{See \url{https://en.wikipedia.org/wiki/King\_Ghidorah}} prior is a mixture of three components -- the first is zero representing practically negligible effect, the second attends to a Gaussian prior for incremental effects, and the third has a Laplace prior watching out for potential breakthrough, heavier-than-Gaussian-tailed, effects. Note that the zero component of Ghidorah is practically important. Our experience conducting many real experiments suggests that for a mature product, in 70\% to 80\% cases, a metric like Revenue may not display any chance of movement and another 10\% could be very weak movement (It is not surprising that we don't have many successful ideas to increase revenue while keeping users happy). For these zero inflated cases, without the Zero component (soft) filtering out these noisy data points, the accuracy of the scale parameters for other components can be significantly hurt.

We chose a mixture prior for its simple posterior mean and variance form. Let $\hat{\mu}_G(\Delta)$ and $\hat{\mu}_L(\Delta)$ be the respective posterior means of the Gaussian and Laplace component, $p_G(\Delta)$ and $p_L(\Delta)$ the respective posterior probabilities of the two components being active (the rest is $0$ component), and $\var_G(\Delta)$ and $\var_L(\Delta)$ the respective posterior variances. We have 
\begin{align}
    \e (\mu|\Delta)  = & p_G(\Delta)\hat{\mu}_G(\Delta) + p_L(\Delta)\hat{\mu}_L(\Delta) \\
    \var (\mu|\Delta)  = & p_G(\Delta)\var_G(\Delta) + p_L(\Delta)\var_L(\Delta) \notag \\
    + & p_G(\Delta)\hat{\mu}_G(\Delta)^2 + p_L(\Delta)\hat{\mu}_L(\Delta)^2 - \e (\mu|\Delta)^2 \label{eq:ghidorahvar}
\end{align}
For each new prediction, we can check the posterior probability of each component and their contribution. When trained on historical data, the larger the prior mixture probability for the Laplace head is, the heavier tail the treatment effect distributes. Similarly, when the zero component has a large weight, the metric is very hard to move.

Fitting the parameters of Ghidorah prior can be done via maximum marginal likelihood, also called MLE-II \citep{barber2012bayesian}. In fact, the model performance could be further improved by incorporating the idea of Stein's unbiased risk estimate (SURE) that tries to optimize an unbiased estimate of the testing error of a model at the same set of predictors $x$ as in the same training set (as if we independently draw another response $y^o$ for those $x$) \citep{efron2004estimation,donoho1995adapting}. 
For example, \citet{xie2012sure} used SURE in an Empirical Bayes setting to improve the prediction accuracy of a James-Stein shrinkage estimator. In the case of Ghidorah, the mixture prior makes it hard to directly use SURE. We use SURE on each individual head to get the initial estimates for the scale parameters and then use Expectation-Maximization (EM) \citep{dempster1977maximum} to iteratively estimate the scale parameters and component weights until convergence.

Similar to the Ghidorah method, the NEST method \citep{fu2020nonparametric} is also an EB method based on the multivariate Tweedie's formula. It models the marginal distribution of the observations $\Delta$ using a non-parametric approach with a two-dimensional Gaussian kernel, which directly extends the \emph{f-modeling} approach proposed in \citep{efron2011tweedie} to solve the heteroskedastic problem. On the contrary, our method, termed as \emph{g-modeling} approach in \citep{efron2011tweedie}, models the prior distribution with a parametric mixture model. The Ghidorah method shows two advantages over the NEST method through our simulation study and empirical analysis. First, the Ghidorah model is more transparent and explainable, which is appealing to industry researchers. Second, non-parametric methods typically require relatively larger number of observations for accurate and numerically stable estimation, as shown in the simulation study in Section 5, which may be a bottleneck for adoption by the A/B testing community.

\subsection{TARwES+: Regression with Ghidorah}
We presented TARwES and Ghidorah as two different methods. In fact, they complement each other. TARwES is a regression framework utilizing EB based predictions as individual features, and Ghidorah predictions can also be used as features here. We call the enhanced method \emph{TARwES+}, showcasing the flexibility of the TARwES framework which can be improved further with better theory-assisted features. Practitioners in a different field might empirically find better mixture priors to be useful when benchmarked on their own data-sets. Another enhancement to TARwES is to go beyond the target metric and include observations or EB predictions of other metrics as features. Consider
$$
(\mu_1,\mu_2)\sim \pi \quad (\Delta_1,\Delta_2)\sim \text{Normal}\big((\mu_1,\mu_2),\Sigma \big) \ .
$$
The information $\Delta_2$ has for $\mu_1$ is greater if 1) the correlation between the underlying movement $\mu_1$ and $\mu_2$ is larger, or 2) the correlation between the noises in $\Sigma$ is lower. In one extreme, $\mu_1=\mu_2$ and noises are uncorrelated. Then $\Delta_2$ contains as much information as $\Delta_1$ has for $\mu_1$. But if the noises are also perfectly correlated, then $\Delta_1=\Delta_2$ and $\Delta_2$ will not be helpful. When considering the practical benefit of adding extra metrics it is crucial to separate the two types of correlations apart. Many metrics with high movement correlation may also have high noise correlations; for example, metrics derived from similar signals.

\section{Confidence interval, Variance reduction and Cold start}\label{sec:extra}

\subsection{Confidence Interval and Adjusted $p-$value}\label{sec:ci}
For the CMLE method, the confidence intervals are computed by inverting the dual hypothesis testing problem \citep{lehmann2006testing,reid2017post}. When selection criteria are one sided, CMLE produces a very asymmetric confidence interval, especially when the observed effect $\Delta$ is close to the selection threshold. For EB method, with posterior mean and variance, a 95\% CI can be computed as 
$$
\e (\mu|\Delta) \pm 1.96 \sqrt{\var (\mu|\Delta)}\ .
$$
This assumes normality of the posterior distribution, which is asymptotically true due to Bernstein-Von Mises theorem \citep{VanderVaart2000}. In this paper, we focus on empirical performance of this CI and take the symmetric form as desired and required. For RwES methods including TARwES, the regression model estimates $\e (\mu|\Delta)$, not $\var (\mu|\Delta)$. But we can use another regression model to estimate $\e (\mu^2|\Delta)$. Let $\Delta_B = \mu + \epsilon_B$ where $\epsilon_B$ is a noise independent of split A and $\mu$,
\begin{align*}
\e (\Delta_B^2|\Delta_A) & = \e(\mu^2|\Delta_A) + 2\e(\mu\epsilon_B|\Delta_A) + \e(\epsilon_B^2|\Delta_A) \\ & = \e(\mu^2|\Delta_A) + \sigma^2/N_B \ .
\end{align*}
Therefore, $\e(\mu^2|\Delta_A) = \e (\Delta_B^2|\Delta_A) - \sigma^2/N_B$ can be used to predict the second moment and the variance. It does require a separate model to be trained and we have to also make sure the predicted variance is numerically stable. For example, regression model might produce a negative variance. EB methods have an advantage over RwES in this regard. Note that, although many regression models also provides a variance for the prediction, it is very different from the posterior variance we need as the two have completely different data generating processes. In practice, we suggest using $\sigma^2/N$ in place of $\var(\mu|\Delta)$, as the latter is usually smaller than $\sigma^2/N$ as explained below. $p-$value can be defined as the smallest $\alpha$ such that the two sided $1-\alpha$ symmetric confidence interval excludes $0$. We define adjusted $p-$value as
\begin{equation}2\times \min\{P(\mu \ge 0|\Delta),P(\mu\le 0|\Delta)\} \ .
\end{equation}\label{eq:adjp}
Although the posterior distribution itself and the Bayesian confidence (or credible) interval contains more information, users familiar with $p-$value can treat this adjusted $p-$value in the same way they use $p-$value to assess significance post selection.

\subsection{Variance Reduction}\label{sec:vr}
Tweedie's formula for the posterior variance \eqref{eq:tweedie} shows 
$$
\var(\mu|X) = \frac{\sigma^2}{N}\{1+ \frac{\sigma^2}{N}\times l''(\Delta)\} \ .
$$
Posterior variance is smaller than $\frac{\sigma^2}{N}$ when $l''(\Delta)<0$. Thus, uncertainty reduction is guaranteed if the marginal likelihood is log-concave, because the marginal likelihood of $\Delta$ is a convolution of the prior density and a Gaussian density of noise. It can be shown that the convolution of two log-concave densities is log-concave. It is easy to see Gaussian is log-concave. Therefore if our prior is log-concave, the empirical Bayes confidence interval will be narrower than the standard unadjusted $\sigma^2/N$. 

Priors like Gaussian, Laplace are log-concave. However, it is not true that the mixture of log-concave distribution is also log-concave. Nevertheless, we found the Ghidorah prior is empirically log-concave for a large range of $\Delta$ that we need to evaluate the posterior variance for. When the variance is not reduced, we found from \eqref{eq:ghidorahvar} that they are cases where the posterior probabilities of the zero and non-zero components are not close to 0 or 1. These are generally uninteresting cases, and it is unclear whether there is a practically significant treatment effect. Practically, we propose to cap the variance by $\sigma^2/N$, so that the confidence interval is always reduced. Our simulation and empirical study showed that this modification still keeps good confidence interval coverage.

\subsection{Cold start: When there is no training data}
EB require a certain number of observations $\Delta_i,i=1,\dots,n$ for parameter estimation. For A/B testing, this typically requires at least 50 to 100 historical experiment data points. This ``cold-start'' problem limits the application of empirical Bayes method when experimenting in a nascent area (a new product or new component to be A/B tested and optimized). Motivated by these real-life scenarios, we propose another method based on local $H_1$ bound \citep{sellke2001calibration}.

We postulate a prior for the effect $\mu$ that is a mixture of $0$ with probability $p$ and a log-concave distribution with probability $1-p$. In statistical null hypothesis testing, the $0$ component is the null hypothesis $H_0$ and the alternative part is the alternative $H_1$. Because of log-concavity, for positive $\Delta$,
$$
\e (\mu|\Delta,H_1)\le \Delta , \quad \var(\mu|\Delta,H_1)\le \sigma^2/N\ .
$$
We can bound the posterior mean by
\begin{equation}
\label{eq:meanub}
\e (\mu|\Delta) 
= P(H_1|\Delta) \e(\mu|\Delta,H_1) \le P(H_1|\Delta) \Delta
\end{equation}
and the posterior variance by 
\begin{align}
\var(\mu|\Delta) 
& = P(H_1|\Delta) \var(\mu|\Delta,H_1) + P(H_1|\Delta)\left\{ 1- P(H_1|\Delta) \right\} \e(\mu|\Delta,H_1)^2 \notag \\
& \le P(H_1|\Delta) \sigma^2 /N + P(H_1|\Delta)^3\left\{ 1-P(H_1|\Delta) \right\}\Delta^2 \label{eq:varub}
\end{align}

\citet{sellke2001calibration} derived a bound for $P(H_1|\Delta)$ when the distribution of $\mu$ under $H_1$ is assumed to be ``local,'' which is a uni-modal distribution centered at 0 with both decaying tails\footnote{See \cite{sellke2001calibration} for the exact technical assumption. This is not so important in our context, as we will be focusing on the empirical performance of the method, and not so much on its theoretical properties.}
\begin{equation}\label{ref:localh1b}
    \frac{P(H_1|\Delta)}{P(H_0|\Delta)}\le \frac{p}{1-p} \times\left (-e z \log (z)\right) \ ,
\end{equation}
where $z$ is the $p$-value of the two-sided hypothesis test. For any given prior odds $p/(1-p)$, \eqref{ref:localh1b} bounds the posterior odds, and hence the posterior $P(H_1|\Delta)$. We then use this to bound both posterior mean and variance in \eqref{eq:meanub} and \eqref{eq:varub}. For negative $\Delta$, we get the lower bound for posterior mean and upper bound for posterior variance. We use these bounds as if they are the posterior mean and variance themselves. Empirical study from real experiment data later in the paper will show that this method works surprisingly well for properly chosen prior odds. 

\section{Simulation Studies}
\label{sec:simulation}

We conduct simulation studies to compare the performances of TARwES and Ghidorah with existing methods -- CMLE, EB with Gaussian, Laplace and Huber prior, NEST, RwES with linear regression and gradient boosted trees.\footnote{R code for implementation of all the mentioned methods as well as simulations are available at \url{https://aka.ms/exp/code/abpostselection}.} We control the ground-truth effect $\mu$ to directly measure prediction accuracy in RMSE and confidence interval coverage. Recall that $\tau^2$ is the prior variance and $\sigma^2/N$ is that of the noise. Therefore $\tau^2N/\sigma^2$ represents the signal-noise-ratio (SNR). To mimic experiments with different scale, we simulate the sample size $N$ from 0.2, 0.5, 1 and 2 millions with equal probability. We conduct three in-depth studies using different priors and they are ordered from simple to hard. For RwES and TARwES, we simply use $\sigma^2/N$ in place of $\var(\mu|\Delta)$ to skip a second regression model for variance for the reason we explained in Section~\ref{sec:ci}. In each study, we first use only 100 training data points (simulated historical experiments), and then increase it to 1000 to see how the performance changes with increased training data. Besides RMSE and confidence interval coverage, we also look at variance reduction/shrikage $Var_S$ by taking the ratio of $\var(\mu|\Delta)$ to the baseline unadjusted MLE's variance $\sigma^2/N$. For CMLE, a z-score threshold of 1.96 was used (comparable to 5\% p-value threshold). We first compute its confidence interval using inverted conditional hypothesis testing, and its width to infer an equivalent variance. For Unadjusted, RwES and TARwES, since we simply took $\sigma^2/N$ as variance we don't need to report variance reduction rate (they are 1). In all cases, we selected SNR to make 10\% to 15\% $p-$values less than 0.1 and 6\% to 8\% less than 0.01, in an attempt to match with the selection rate we observed in real experiments. All RMSE numbers are scaled with the unit of $0.1$ for easier comparison. All SNR numbers are computed with 1 million sample size (SNR = 1E6$\times\tau^2/\sigma^2$). In each study, we also implemented the theoretical conditional mean and variance to show or approximate the best possible results assuming we do know the true oracle prior distribution.

\subsection{Case 1: Gaussian Prior with SNR 0.1}

Results in Table~\ref{tab:sim1}. Gaussian prior is the simplest and we expect many existing methods such as J-S shrinkage to do well. In this setting, we saw about 15.2\% cases passed $p-$value 5\% threshold and 6.5\% passed 1\%. Our main observations are the following. First, all methods improved upon unadjusted prediction. Second, CMLE showed higher RMSE. Its confidence interval coverage is 100\% but with variance ratio of 4 to 6. This means the high coverage is at the cost of much wider intervals. As expected, CMLE confidence interval can be wildly asymmetric when the cutoff threshold is closer to the observed $\Delta$. Numerical stability of CMLE confidence interval remains an issue. Based on its poor RMSE, we did not seek to further improve its stability. Third, not surprisingly, EB with Gaussian prior showed good performance, and with 1000 training data it is very close to the theoretical best. But Ghidorah performed almost the same, as did Huber prior. Their performance for 100 training points is already close to EB with Gaussian prior. RwES methods weren't as good as EB methods. GBT's performance improved significantly with increased training data. TARwES significantly improved upon RwES, and was at par with Ghidorah and Gaussian EB for 1000 training data. Lastly, NEST method showed the worst RMSE and yielded slightly undercovered confidence interval in general but the performance improves with larger training set, which indicates that a non-parametric method does require more samples to perform accurate inference.
\begin{table}[hbt]
\caption{Gaussian Prior}
\label{tab:sim1}
\resizebox{0.48\textwidth}{!}{%
\begin{tabular}{llclllllll}
{\ul } & \multicolumn{2}{l}{Case 1: Gaussian  Prior(SNR=0.1) }     &                                      & \multicolumn{3}{c}{100 Training Data Points}        & \multicolumn{3}{c}{1000 Training Data Points} \\ \cline{2-10} 
       & Method        & \multicolumn{1}{c|}{Selected} & \multicolumn{1}{l|}{$p-$val}         & RMSE & Coverage & \multicolumn{1}{l|}{$Var_S$} & RMSE  & Coverage & $Var_S$               \\ \cline{2-10} 
       & Unadjusted    & \multicolumn{1}{l|}{6.5\%}    & \multicolumn{1}{l|}{\textless{}0.01} & 2.17      & 70.6\%   & \multicolumn{1}{c|}{-}       & 2.17       & 70.6\%   & \multicolumn{1}{c}{-} \\
       &               & \multicolumn{1}{l|}{15.2\%}   & \multicolumn{1}{l|}{\textless{}0.05} & 2.16      & 78.1\%   & \multicolumn{1}{c|}{-}       & 2.16       & 78.1\%   & \multicolumn{1}{c}{-} \\
       &               & \multicolumn{1}{l|}{100\%}    & \multicolumn{1}{l|}{All}             & 1.46      & 95.0\%   & \multicolumn{1}{c|}{-}       & 1.46       & 95.0\%   & \multicolumn{1}{c}{-} \\ \cline{2-10} 
       & Theoretical   & \multicolumn{1}{l|}{6.5\%}    & \multicolumn{1}{l|}{\textless{}0.01} & 0.68      & 94.5\%   & \multicolumn{1}{l|}{0.54}    & 0.68       & 94.5\%   & 0.54                  \\
       & (known prior) & \multicolumn{1}{l|}{15.2\%}   & \multicolumn{1}{l|}{\textless{}0.05} & 0.71      & 94.7\%   & \multicolumn{1}{l|}{0.50}    & 0.71       & 94.7\%   & 0.50                  \\
       &               & \multicolumn{1}{l|}{100\%}    & \multicolumn{1}{l|}{All}             & 0.77      & 94.9\%   & \multicolumn{1}{l|}{0.41}    & 0.77       & 94.9\%   & 0.41                  \\ \hline
       & CMLE          & \multicolumn{1}{l|}{6.5\%}    & \multicolumn{1}{l|}{\textless{}0.01} &  1.69         &  100\%       & \multicolumn{1}{l|}{5.93}        &    1.69        &   100\%       &         5.93              \\
       &              & \multicolumn{1}{l|}{15.2\%}   & \multicolumn{1}{l|}{\textless{}0.05} & 1.30          &    100\%      & \multicolumn{1}{l|}{5.53}        &   1.30         &     100\%     &        5.53               \\
       &               & \multicolumn{1}{l|}{100\%}    & \multicolumn{1}{l|}{All}             &   1.06        &  100\%        & \multicolumn{1}{l|}{4.17}        &   1.06         &   100\%       &        4.17               \\ \hline
       & Gaussian        & \multicolumn{1}{l|}{6.5\%}    & \multicolumn{1}{l|}{\textless{}0.01} & 0.76      & 90.1\%   & \multicolumn{1}{l|}{0.52}    & 0.70       & 94.0\%   & 0.54                  \\
       &               & \multicolumn{1}{l|}{15.2\%}   & \multicolumn{1}{l|}{\textless{}0.05} & 0.77      & 90.9\%   & \multicolumn{1}{l|}{0.48}    & 0.72       & 94.4\%   & 0.50                  \\
       &               & \multicolumn{1}{l|}{100\%}    & \multicolumn{1}{l|}{All}             & 0.78      & 92.6\%   & \multicolumn{1}{l|}{0.40}    & 0.77       & 94.7\%   & 0.42                  \\ \cline{2-10} 
       & Laplace       & \multicolumn{1}{l|}{6.5\%}    & \multicolumn{1}{l|}{\textless{}0.01} & 0.77      & 97.1\%   & \multicolumn{1}{l|}{0.83}    & 0.78       & 97.7\%   & 0.85                  \\
       &               & \multicolumn{1}{l|}{15.2\%}   & \multicolumn{1}{l|}{\textless{}0.05} & 0.77      & 96.0\%   & \multicolumn{1}{l|}{0.70}    & 0.74       & 97.7\%   & 0.73                  \\
       &               & \multicolumn{1}{l|}{100\%}    & \multicolumn{1}{c|}{All}             & 0.78      & 91.8\%   & \multicolumn{1}{l|}{0.39}    & 0.74       & 93.5\%   & 0.85                  \\ \cline{2-10} 
       & Huber         & \multicolumn{1}{l|}{6.5\%}    & \multicolumn{1}{c|}{\textless{}0.01} & 0.71      & 92.8\%   & \multicolumn{1}{l|}{0.52}    & 0.69       & 94.4\%   & 0.54                  \\
       &               & \multicolumn{1}{l|}{15.2\%}   & \multicolumn{1}{l|}{\textless{}0.05} & 0.73      & 92.9\%   & \multicolumn{1}{l|}{0.48}    & 0.72       & 94.6\%   & 0.50                  \\
       &               & \multicolumn{1}{l|}{100\%}    & \multicolumn{1}{c|}{All}             & 0.77      & 93.7\%   & \multicolumn{1}{l|}{0.40}    & 0.77       & 94.8\%   & 0.42                  \\ \cline{2-10} 
       & Ghidorah      & \multicolumn{1}{l|}{6.5\%}    & \multicolumn{1}{l|}{\textless{}0.01} & 0.76      & 94.3\%   & \multicolumn{1}{l|}{0.69}    & 0.71       & 95.7\%   & 0.66                  \\
       &               & \multicolumn{1}{l|}{15.2\%}   & \multicolumn{1}{l|}{\textless{}0.05} & 0.77      & 93.9\%   & \multicolumn{1}{l|}{0.62}    & 0.73       & 96.0\%   & 0.61                  \\
       &               & \multicolumn{1}{l|}{100\%}    & \multicolumn{1}{l|}{All}             & 0.78      & 91.0\%   & \multicolumn{1}{l|}{0.38}    & 0.77       & 94.2\%   & 0.41                  \\ \hline
       & RwES(Linear)  & \multicolumn{1}{l|}{6.5\%}    & \multicolumn{1}{l|}{\textless{}0.01} & 1.07      & 87.3\%   & \multicolumn{1}{c|}{-}       & 1.01       & 89.8\%   & \multicolumn{1}{c}{-} \\
       &               & \multicolumn{1}{l|}{15.2\%}   & \multicolumn{1}{l|}{\textless{}0.05} & 1.00      & 92.1\%   & \multicolumn{1}{c|}{-}       & 0.94       & 94.0\%   & \multicolumn{1}{c}{-} \\
       &               & \multicolumn{1}{l|}{100\%}    & \multicolumn{1}{l|}{All}             & 0.86      & 97.9\%   & \multicolumn{1}{c|}{-}       & 0.83       & 98.4\%   & \multicolumn{1}{c}{-} \\ \cline{2-10} 
       & RwES(GBT)     & \multicolumn{1}{l|}{6.5\%}    & \multicolumn{1}{l|}{\textless{}0.01} & 1.22      & 81.9\%   & \multicolumn{1}{c|}{-}       & 0.95       & 91.5\%   & \multicolumn{1}{c}{-} \\
       &               & \multicolumn{1}{l|}{15.2\%}   & \multicolumn{1}{l|}{\textless{}0.05} & 1.07      & 89.7\%   & \multicolumn{1}{c|}{-}       & 0.88       & 95.1\%   & \multicolumn{1}{c}{-} \\
       &               & \multicolumn{1}{l|}{100\%}    & \multicolumn{1}{l|}{All}             & 0.91      & 97.0\%   & \multicolumn{1}{c|}{-}       & 0.82       & 98.5\%   & \multicolumn{1}{c}{-} \\ \cline{2-10} 
       & TARwES        & \multicolumn{1}{l|}{6.5\%}    & \multicolumn{1}{l|}{\textless{}0.01} & 0.83      & 95.5\%   & \multicolumn{1}{c|}{-}       & 0.70       & 98.5\%   & \multicolumn{1}{c}{-} \\
       &               & \multicolumn{1}{l|}{15.2\%}   & \multicolumn{1}{l|}{\textless{}0.05} & 0.81      & 97.2\%   & \multicolumn{1}{c|}{-}       & 0.72       & 99.0\%   & \multicolumn{1}{c}{-} \\
       &               & \multicolumn{1}{l|}{100\%}    & \multicolumn{1}{l|}{All}             & 0.79      & 99.1\%   & \multicolumn{1}{c|}{-}       & 0.77       & 99.4\%   & \multicolumn{1}{c}{-} \\ \cline{2-10}
       & NEST        & \multicolumn{1}{l|}{6.5\%}    & \multicolumn{1}{l|}{\textless{}0.01} & 2.05      & 89.4\%   & \multicolumn{1}{c|}{0.62}       & 1.15       & 91.9\%   & \multicolumn{1}{c}{0.51} \\
       &               & \multicolumn{1}{l|}{15.2\%}   & \multicolumn{1}{l|}{\textless{}0.05} & 1.90      & 91.5\%   & \multicolumn{1}{c|}{0.55}       & 1.04       & 92.7\%   & \multicolumn{1}{c}{0.59} \\
       &               & \multicolumn{1}{l|}{100\%}    & \multicolumn{1}{l|}{All}             & 1.32      & 94.1\%   & \multicolumn{1}{c|}{0.48}       & 0.89       & 93.4\%   & \multicolumn{1}{c}{0.49} \\
       
\end{tabular}%
}
\end{table}

\subsection{Case 2: 50\% Zero and 50\% T-prior with degrees of freedom 3 and SNR 0.4} 
Results in Table~\ref{tab:sim2}. With 50\% 0 effect and 50\% heavy tailed t-distribution, 14.5\% passed $p-$value threshold 5\% and 6.5\% less than 1\%. This case is harder than Gaussian prior and could create issues for Gaussian prior EB. With only 100 training data, only 50 data points are effective for estimating prior scale parameter, and the other 50 are adding noise. We found that RwES with linear and GBT, and NEST struggled even with 1000 training data --- they are not much better than Unadjusted, and GBT and NEST are worse than linear regression. Ghidorah gave the best results, followed by Laplace prior. At 100 training data, Ghidorah already beat the approximated theoretical best (using first order approximation for t-prior as exact formula does not exist \citep{pericchi1992exact}) in both RMSE and coverage. Increasing training data further improved RMSE and interval coverage. Both Gaussian and Huber prior showed mediocre performance, as does CMLE. TARwES is only slightly worse than Ghidorah with 1000 training data.   

\begin{table}[hbt]
\caption{Mixture of 50\% Zero and 50\% T }
\label{tab:sim2}
\resizebox{0.48\textwidth}{!}{%
\begin{tabular}{llllllllll}
{\ul } & \multicolumn{3}{l}{Case 2: 50\% Zero/50\%T-prior(df=3,SNR=0.4).}                                                & \multicolumn{3}{c}{100 Training Data Points}        & \multicolumn{3}{c}{1000 Training Data Points} \\ \cline{2-10} 
       & Method              & \multicolumn{1}{c|}{Selected} & \multicolumn{1}{l|}{$p-$val}         & RMSE & Coverage & \multicolumn{1}{l|}{$Var_S$} & RMSE  & Coverage & $Var_S$               \\ \cline{2-10} 
       & Unadjusted          & \multicolumn{1}{l|}{7.6\%}    & \multicolumn{1}{l|}{\textless{}0.01} & 1.95      & 79.5\%   & \multicolumn{1}{c|}{-}       & 1.99       & 79.5\%   & \multicolumn{1}{c}{-} \\
       &                     & \multicolumn{1}{l|}{14.5\%}   & \multicolumn{1}{l|}{\textless{}0.05} & 2.17      & 71.6\%   & \multicolumn{1}{c|}{-}       & 2.18       & 71.6\%   & \multicolumn{1}{c}{-} \\
       &                     & \multicolumn{1}{l|}{100\%}    & \multicolumn{1}{l|}{All}             & 1.45      & 95.0\%   & \multicolumn{1}{c|}{-}       & 1.46       & 95.0\%   & \multicolumn{1}{c}{-} \\ \cline{2-10} 
       & Theoretical(approx) & \multicolumn{1}{l|}{7.6\%}    & \multicolumn{1}{l|}{\textless{}0.01} & 1.45      & 88.3\%   & \multicolumn{1}{c|}{-}       & 1.45       & 88.3\%   & \multicolumn{1}{c}{-} \\
       & (Known Prior)       & \multicolumn{1}{l|}{14.5\%}   & \multicolumn{1}{l|}{\textless{}0.05} & 1.45      & 88.2\%   & \multicolumn{1}{c|}{-}       & 1.45       & 88.2\%   & \multicolumn{1}{c}{-} \\
       &                     & \multicolumn{1}{l|}{100\%}    & \multicolumn{1}{l|}{All}             & 0.90      & 96.9\%   & \multicolumn{1}{c|}{-}       & 0.90       & 96.9\%   & \multicolumn{1}{c}{-} \\ \hline
       & CMLE                & \multicolumn{1}{l|}{7.6\%}    & \multicolumn{1}{l|}{\textless{}0.01} &   1.63        &   100\%       & \multicolumn{1}{l|}{5.46}        &    1.63        &   100\%       &     5.46                  \\
       &                     & \multicolumn{1}{l|}{14.5\%}   &  \multicolumn{1}{l|}{\textless{}0.05} &     1.45     &   100\%       & \multicolumn{1}{l|}{5.34}        &      1.45      &     100\%     &      5.34                 \\
       &                     & \multicolumn{1}{l|}{100\%}    & \multicolumn{1}{l|}{All}             &   1.08        &    100\%      & \multicolumn{1}{l|}{4.11}        &   1.08         & 100\%         &      4.11                 \\ \hline
       & Gaussian              & \multicolumn{1}{l|}{7.6\%}    & \multicolumn{1}{l|}{\textless{}0.01} & 1.91      & 71.3\%   & \multicolumn{1}{l|}{0.61}    & 1.80       & 75.9\%   & 0.64                  \\
       &                     & \multicolumn{1}{l|}{14.5\%}   & \multicolumn{1}{l|}{\textless{}0.05} & 1.65      & 78.5\%   & \multicolumn{1}{l|}{0.58}    & 1.55       & 82.1\%   & 0.61                  \\
       &                     & \multicolumn{1}{l|}{100\%}    & \multicolumn{1}{l|}{All}             & 0.96      & 93.9\%   & \multicolumn{1}{l|}{0.52}    & 0.93       & 95.2\%   & 0.55                  \\ \cline{2-10} 
       & Laplace             & \multicolumn{1}{l|}{7.6\%}    & \multicolumn{1}{l|}{\textless{}0.01} & 1.37      & 89.2\%   & \multicolumn{1}{l|}{0.91}    & 1.33       & 90.9\%   & 0.93                  \\
       &                     & \multicolumn{1}{l|}{14.5\%}   & \multicolumn{1}{l|}{\textless{}0.05} & 1.30      & 91.1\%   & \multicolumn{1}{l|}{0.80}    & 1.26       & 92.9\%   & 0.83                  \\
       &                     & \multicolumn{1}{l|}{100\%}    & \multicolumn{1}{l|}{All}             & 0.85      & 95.0\%   & \multicolumn{1}{l|}{0.48}    & 0.83       & 96.1\%   & 0.50                  \\ \cline{2-10} 
       & Huber               & \multicolumn{1}{l|}{7.6\%}    & \multicolumn{1}{l|}{\textless{}0.01} & 1.85     & 70.8\%   & \multicolumn{1}{l|}{0.61}    & 1.78       & 76.0\%   & 0.64                  \\
       &                     & \multicolumn{1}{l|}{14.5\%}   & \multicolumn{1}{l|}{\textless{}0.05} & 1.59      & 78.4\%   & \multicolumn{1}{l|}{0.57}    & 1.54       & 82.3\%   & 0.61                  \\
       &                     & \multicolumn{1}{l|}{100\%}    & \multicolumn{1}{l|}{All}             & 0.94      & 93.7\%   & \multicolumn{1}{l|}{0.50}    & 0.92       & 95.2\%   & 0.54                  \\ \cline{2-10} 
       & Ghidorah            & \multicolumn{1}{l|}{7.6\%}    & \multicolumn{1}{l|}{\textless{}0.01} & 1.34      & 89.5\%   & \multicolumn{1}{l|}{0.94}    & 1.26       & 93.1\%   & 0.98                  \\
       &                     & \multicolumn{1}{l|}{14.5\%}   & \multicolumn{1}{l|}{\textless{}0.05} & 1.28      & 91.0\%   & \multicolumn{1}{l|}{0.86}    & 1.20       & 94.0\%   & 0.91                  \\
       &                     & \multicolumn{1}{l|}{100\%}    & \multicolumn{1}{l|}{All}             & 0.82      & 91.2\%   & \multicolumn{1}{l|}{0.36}    & 0.80       & 92.8\%   & 0.38                  \\ \hline
       & RwES(Linear)        & \multicolumn{1}{l|}{7.6\%}    & \multicolumn{1}{l|}{\textless{}0.01} & 2.33      & 68.3\%   & \multicolumn{1}{c|}{-}       & 2.18       & 71.6\%   & \multicolumn{1}{c}{-} \\
       &                     & \multicolumn{1}{l|}{14.5\%}   & \multicolumn{1}{l|}{\textless{}0.05} & 1.94      & 80.8\%   & \multicolumn{1}{c|}{-}       & 1.82       & 83.0\%   & \multicolumn{1}{c}{-} \\
       &                     & \multicolumn{1}{l|}{100\%}    & \multicolumn{1}{l|}{All}             & 1.07      & 96.5\%   & \multicolumn{1}{c|}{-}       & 1.02       & 97.0\%   & \multicolumn{1}{c}{-} \\ \cline{2-10} 
       & RwES(GBT)           & \multicolumn{1}{l|}{7.6\%}    & \multicolumn{1}{l|}{\textless{}0.01} & 3.49      & 51.5\%   & \multicolumn{1}{c|}{-}       & 2.71       & 67.6\%   & \multicolumn{1}{c}{-} \\
       &                     & \multicolumn{1}{l|}{14.5\%}   & \multicolumn{1}{l|}{\textless{}0.05} & 2.70      & 69.9\%   & \multicolumn{1}{c|}{-}       & 2.23       & 79.2\%   & \multicolumn{1}{c}{-} \\
       &                     & \multicolumn{1}{l|}{100\%}    & \multicolumn{1}{l|}{All}             & 1.31      & 94.3\%   & \multicolumn{1}{c|}{-}       & 1.11       & 96.1\%   & \multicolumn{1}{c}{-} \\ \cline{2-10} 
       & TARwES              & \multicolumn{1}{l|}{7.6\%}    & \multicolumn{1}{l|}{\textless{}0.01} & 1.56      & 85.6\%   & \multicolumn{1}{c|}{-}       & 1.36       & 91.1\%   & \multicolumn{1}{c}{-} \\
       &                     & \multicolumn{1}{l|}{14.5\%}   & \multicolumn{1}{l|}{\textless{}0.05} & 1.42      & 90.4\%   & \multicolumn{1}{c|}{-}       & 1.27       & 94.1\%   & \multicolumn{1}{c}{-} \\
       &                     & \multicolumn{1}{l|}{100\%}    & \multicolumn{1}{l|}{All}             & 0.88      & 98.1\%   & \multicolumn{1}{c|}{-}       & 0.83       & 98.6\%   & \multicolumn{1}{c}{-} \\ \cline{2-10} 
       & NEST              & \multicolumn{1}{l|}{7.6\%}    & \multicolumn{1}{l|}{\textless{}0.01} & 2.59      & 84.2\%   & \multicolumn{1}{c|}{0.77}       & 2.47       & 88.8\%   & \multicolumn{1}{c}{0.88} \\
       &                     & \multicolumn{1}{l|}{14.5\%}   & \multicolumn{1}{l|}{\textless{}0.05} & 2.26      & 86.7\%   & \multicolumn{1}{c|}{0.78}       & 2.12       & 91.7\%   & \multicolumn{1}{c}{0.87} \\
       &                     & \multicolumn{1}{l|}{100\%}    & \multicolumn{1}{l|}{All}             & 1.29      & 94.8\%   & \multicolumn{1}{c|}{0.52}       & 1.14       & 95.1\%   & \multicolumn{1}{c}{0.51}
\end{tabular}%
}
\end{table}

\begin{table}[hbt]
\caption{Mixture of 90\% Zero and 10\% T-prior}
\label{tab:sim3}
\resizebox{0.48\textwidth}{!}{%
\begin{tabular}{llllllllll}
{\ul } & \multicolumn{3}{l}{Case 3: 90\% Zero/10\% T-prior(df=3,SNR=10)}                                      & \multicolumn{3}{c}{100 Training Data Points}   & \multicolumn{3}{c}{1000 Training Data Points} \\ \cline{2-10} 
       & Method              & \multicolumn{1}{c|}{Selected} & \multicolumn{1}{l|}{$p-$val}         & RMSE & Coverage & \multicolumn{1}{l|}{$Var_S$} & RMSE   & Coverage   & $Var_S$                 \\ \cline{2-10} 
       & Unadjusted          & \multicolumn{1}{l|}{7.14\%}   & \multicolumn{1}{l|}{\textless{}0.01} & 1.94 & 82.7\%   & \multicolumn{1}{c|}{-}       & 1.94   & 82.7\%     & \multicolumn{1}{c}{-}   \\
       &                     & \multicolumn{1}{l|}{11.6\%}   & \multicolumn{1}{l|}{\textless{}0.05} & 2.40 & 57.5\%   & \multicolumn{1}{c|}{-}       & 2.40   & 57.5\%     & \multicolumn{1}{c}{-}   \\
       &                     & \multicolumn{1}{l|}{100\%}    & \multicolumn{1}{l|}{All}             & 1.46 & 94.9\%   & \multicolumn{1}{c|}{-}       & 1.46   & 94.9\%     & \multicolumn{1}{c}{-}   \\ \cline{2-10} 
       & Theoretical(approx) & \multicolumn{1}{l|}{7.14\%}   & \multicolumn{1}{l|}{\textless{}0.01} & 1.61 & 90.4\%   & \multicolumn{1}{c|}{-}       & 1.61   & 90.4\%     & \multicolumn{1}{c}{-}   \\
       & (Known Prior)       & \multicolumn{1}{l|}{11.6\%}   & \multicolumn{1}{l|}{\textless{}0.05} & 1.53 & 91.2\%   & \multicolumn{1}{c|}{-}       & 1.53   & 91.2\%     & \multicolumn{1}{c}{-}   \\
       &                     & \multicolumn{1}{l|}{100\%}    & \multicolumn{1}{l|}{All}             & 0.65 & 98.5\%   & \multicolumn{1}{c|}{-}       & 0.65   & 98.5\%     & \multicolumn{1}{c}{-}   \\ \hline
       & CMLE                & \multicolumn{1}{l|}{7.14\%}   & \multicolumn{1}{l|}{\textless{}0.01} &   1.69   &      100\%    & \multicolumn{1}{l|}{4.75}        &   1.69     &     100\%       &    4.75                     \\
       &                     & \multicolumn{1}{l|}{11.6\%}   &  \multicolumn{1}{l|}{\textless{}0.05} & 1.60     &    100\%      & \multicolumn{1}{l|}{4.92}        &    1.60    &      100\%      &        4.92                 \\
       &                     & \multicolumn{1}{l|}{100\%}    & \multicolumn{1}{l|}{All}             &   1.04   &     100\%     & \multicolumn{1}{l|}{4.00}        &   1.04     &         100\%   &    4.00                     \\ \hline
       & Gaussian              & \multicolumn{1}{l|}{7.14\%}   & \multicolumn{1}{l|}{\textless{}0.01} & 3.73 & 54.4\%   & \multicolumn{1}{l|}{0.75}    & 2.90   & 65.3\%     & 0.83                    \\
       &                     & \multicolumn{1}{l|}{11.6\%}   & \multicolumn{1}{l|}{\textless{}0.05} & 3.25 & 58.1\%   & \multicolumn{1}{l|}{0.74}    & 2.70   & 61.6\%     & 0.82                    \\
       &                     & \multicolumn{1}{l|}{100\%}    & \multicolumn{1}{l|}{All}             & 1.39 & 94.8\%   & \multicolumn{1}{l|}{0.73}    & 1.28   & 95.3\%     & 0.80                    \\ \cline{2-10} 
       & Laplace             & \multicolumn{1}{l|}{7.14\%}   & \multicolumn{1}{l|}{\textless{}0.01} & 1.97 & 81.0\%   & \multicolumn{1}{l|}{0.97}    & 1.78   & 83.2\%     & 0.99                    \\
       &                     & \multicolumn{1}{l|}{11.6\%}   & \multicolumn{1}{l|}{\textless{}0.05} & 1.96 & 84.1\%   & \multicolumn{1}{l|}{0.91}    & 1.89   & 84.7\%     & 0.95                    \\
       &                     & \multicolumn{1}{l|}{100\%}    & \multicolumn{1}{l|}{All}             & 0.95 & 97.7\%   & \multicolumn{1}{l|}{0.64}    & 0.98   & 97.9\%     & 0.70                    \\ \cline{2-10} 
       & Huber               & \multicolumn{1}{l|}{7.14\%}   & \multicolumn{1}{l|}{\textless{}0.01} & 3.34 & 57.3\%   & \multicolumn{1}{l|}{0.78}    & 2.77   & 66.6\%     & 0.84                    \\
       &                     & \multicolumn{1}{l|}{11.6\%}   & \multicolumn{1}{l|}{\textless{}0.05} & 2.98 & 59.5\%   & \multicolumn{1}{l|}{0.76}    & 2.60   & 63.0\%     & 0.82                    \\
       &                     & \multicolumn{1}{l|}{100\%}    & \multicolumn{1}{l|}{All}             & 1.32 & 95.0\%   & \multicolumn{1}{l|}{0.74}    & 1.25   & 95.4\%     & 0.80                    \\ \cline{2-10} 
       & Ghidorah            & \multicolumn{1}{l|}{7.6\%}    & \multicolumn{1}{l|}{\textless{}0.01} & 1.64 & 90.6\%   & \multicolumn{1}{l|}{0.99}    & 1.60   & 91.3\%     & 1.00                    \\
       &                     & \multicolumn{1}{l|}{14.5\%}   & \multicolumn{1}{l|}{\textless{}0.05} & 1.56 & 91.1\%   & \multicolumn{1}{l|}{0.91}    & 1.52   & 91.7\%     & 0.90                    \\
       &                     & \multicolumn{1}{l|}{100\%}    & \multicolumn{1}{l|}{All}             & 0.65 & 97.11\%  & \multicolumn{1}{l|}{0.20}    & 0.64   & 97.1\%     & 0.19                    \\ \hline
       & RwES(Linear)        & \multicolumn{1}{l|}{7.14\%}   & \multicolumn{1}{l|}{\textless{}0.01} & 3.80 & 54.8\%   & \multicolumn{1}{c|}{-}       & 2.90   & 60.0\%     & \multicolumn{1}{c}{-}   \\
       &                     & \multicolumn{1}{l|}{11.6\%}   & \multicolumn{1}{l|}{\textless{}0.05} & 3.35 & 66.5\%   & \multicolumn{1}{c|}{-}       & 2.73   & 70.9\%     & \multicolumn{1}{c}{-}   \\
       &                     & \multicolumn{1}{l|}{100\%}    & \multicolumn{1}{l|}{All}             & 1.48 & 95.9\%   & \multicolumn{1}{c|}{-}       & 1.34   & 96.5\%     & \multicolumn{1}{c}{-}   \\ \cline{2-10} 
       & RwES(GBT)           & \multicolumn{1}{l|}{7.14\%}   & \multicolumn{1}{l|}{\textless{}0.01} & 10.8 & 22.4\%   & \multicolumn{1}{c|}{-}       & 7.24   & 45.1\%     & \multicolumn{1}{c}{-}   \\
       &                     & \multicolumn{1}{l|}{11.6\%}   & \multicolumn{1}{l|}{\textless{}0.05} & 8.52 & 45.5\%   & \multicolumn{1}{c|}{-}       & 5.85   & 61.6\%     & \multicolumn{1}{c}{-}   \\
       &                     & \multicolumn{1}{l|}{100\%}    & \multicolumn{1}{l|}{All}             & 3.04 & 91.0\%   & \multicolumn{1}{c|}{-}       & 2.04   & 95.1\%     & \multicolumn{1}{c}{-}   \\ \cline{2-10} 
       & TARwES              & \multicolumn{1}{l|}{7.14\%}   & \multicolumn{1}{l|}{\textless{}0.01} & 2.47 & 72.7\%   & \multicolumn{1}{c|}{-}       & 1.97   & 79.1\%     & \multicolumn{1}{c}{-}   \\
       &                     & \multicolumn{1}{l|}{11.6\%}   & \multicolumn{1}{l|}{\textless{}0.05} & 2.29 & 80.4\%   & \multicolumn{1}{c|}{-}       & 1.97   & 85.0\%     & \multicolumn{1}{c}{-}   \\
       &                     & \multicolumn{1}{l|}{100\%}    & \multicolumn{1}{l|}{All}             & 1.02 & 97.6\%   & \multicolumn{1}{c|}{-}       & 0.98   & 98.1\%     & \multicolumn{1}{c}{-}  \\ \hline
       & TARwES+              & \multicolumn{1}{l|}{7.14\%}   & \multicolumn{1}{l|}{\textless{}0.01} & 2.09 & 81.9\%   & \multicolumn{1}{c|}{-}       & 1.62   & 90.5\%     & \multicolumn{1}{c}{-}   \\
       &                     & \multicolumn{1}{l|}{11.6\%}   & \multicolumn{1}{l|}{\textless{}0.05} & 1.89 & 86.5\%   & \multicolumn{1}{c|}{-}       & 1.54   & 91.8\%     & \multicolumn{1}{c}{-}   \\
       &                     & \multicolumn{1}{l|}{100\%}    & \multicolumn{1}{l|}{All}             & 0.75 & 98.1\%   & \multicolumn{1}{c|}{-}       & 0.65   & 98.6\%     & \multicolumn{1}{c}{-} \\ \hline
       & NEST              & \multicolumn{1}{l|}{7.14\%}   & \multicolumn{1}{l|}{\textless{}0.01} & 3.29 & 76.8\%   & \multicolumn{1}{c|}{0.86}       & 2.57   & 84.5\%     & \multicolumn{1}{c}{1.03}   \\
       &                     & \multicolumn{1}{l|}{11.6\%}   & \multicolumn{1}{l|}{\textless{}0.05} & 2.97 & 79.5\%   & \multicolumn{1}{c|}{0.88}       & 2.39   & 87.2\%     & \multicolumn{1}{c}{1.00}   \\
       &                     & \multicolumn{1}{l|}{100\%}    & \multicolumn{1}{l|}{All}             & 1.26 & 96.6\%   & \multicolumn{1}{c|}{0.56}       & 1.02   & 97.7\%     & \multicolumn{1}{c}{0.55} 
\end{tabular}%
}
\end{table}

\subsection{Case 3: 90\% Zero and 10\% T-prior with degrees of freedom 3 and SNR 10} 
Results in Table~\ref{tab:sim3}. This is the hardest case. With 100 training data, only 10 points are effective for prior parameter estimation, with 90 points adding noise. This prior represents no effect or break-through. This can be seen from the fact that among 11.6\% with $p-$value less than 5\%, 7.14\% are less than 1\%. RMSE of Unadjusted prediction post-selection of $p-$value<0.01 wasn't too bad --- very little adjustment is needed at tail. Similar to the last case, two RwES methods (especially GBT), NEST, and EB with Gaussian and Huber prior didn't do well. Ghidorah still performed the best, very close to the approximated theoretical best. Laplace prior was better than Gaussian and Huber prior, but wasn't as good as Ghidorah, with a big gap. We think the reason is that all EB priors without modeling zero component did poorly in estimating the scale parameter because they cannot remove 90\% noisy data points in this process. As a result, TARwES also didn't perform well similar to its theory-assisted features. We added Ghidorah prediction as an additional feature in TARwES to make TARwES+. It greatly improved TARwES and was close to Ghidorah with 1000 training data. Even though it contains Ghidorah prediction as a feature, at 100 training data it performed worse than Ghidorah since the extra layer of regression increases model complexity and has slightly higher variance.



\section{Real Experiments Data Studies}
\label{sec:empstudy}

We evaluated and compared our new methods to existing methods using a real experiment data-set from a business unit of a large product with millions of active users. We did a thorough data quality check and experiments with known trustworthy issues such as sample ratio mismatch \citep{fabijan2019diagnosing} were filtered out. All experiments ran for 1 week. We further only included experiments with at least 1m sample size and the remaining sample size ranges from 1m to more than 50m. More than 1000 experiments were used in this study, from which we randomly put half as the training set and the rest as the test set. For experiments in both sets, we further split them into two. This is required for testing as we will apply various methods to split A and compare it with the ``ground-truth'' observed in split B. For training, EB based methods like Ghidorah do not need splitting, while RwES and TARwES methods train on split data. 

We looked at two top line metrics, one measuring user engagement and the other, site performance (page-loading-time). Table~\ref{tab:empirical} shows the results on the test set. For the engagement metric, 13.1\% (68/519) $p-$values were less than 5\% and 6.6\%(34/519) less than 1\%; the site performance metric had 14.4\%(75/522) less than 5\% and 7.7\%(40/522) less than 1\%. These numbers are common for a mature product that has been under heavy optimization using A/B testing. We included TARwES+, which adds Ghidorah prediction as one of the features in addition to Gaussian and Laplace priors. We also included Local $H_1$ method we proposed that can be used without any training data. We used a naive prior odds of 1:1 and also a rough expert's guess version where we use 1:7 for engagement metric and 1:6 for site performance metric to make the prior probability of $H_1$ close to the proportion of $p-$values less than 5\%. We also included LaplaceFitGhidorah, a Laplace distribution (which is log-concave, as mentioned earlier) fit to Ghidorah, by minimizing KL-divergence with an approximation to Ghidorah. The Ghidorah approximation comprised a low variance Normal distribution, instead of the zero component of Ghidorah, in addition to the Normal and Laplace components. We removed RwES with GBT from the contestants given inferior results from simulation studies. 

We drew the following conclusions. Most methods helped to reduce RMSE and improved coverage, except RwES with linear regression. This is partly due to a large range of different sample sizes, unlike the situation of all similar sample sizes reported in \citep{coey2019improving}. Ghidorah did very well and was the best in both accuracy and coverage, while significantly reduced confidence interval width when $p-$values are not small. Huber prior and LaplaceFitGhidorah closely followed Ghidorah for both metrics. Laplace and Gaussian prior showed a clear gap to Ghidorah, Huber, and LaplaceFitGhidorah. TARwES uses Gaussian and Laplace EB as its features so its performance is similar to the two. TARwES+ added Ghidorah as a new feature. It also did very well and was close to Ghidorah. For local $H_1$ methods, the performance depends heavily on the prior odds. 1:1 prior odds clearly wasn't good. With a bit domain knowledge tuning, 1:7 and 1:6 priors for the two metrics showed much better results, albeit still not as good as Ghidorah and TARwES+. Nevertheless, the fact that local $H_1$ method with a rough expert's guess can already greatly improve accuracy of treatment effect prediction over unadjusted and even achieve similar performance of James-Stein (Gaussian prior) is satisfying. It is extremely easy to apply in practice. NEST in general performs poorly, partially due to the fact that the training sample size is not large enough to estimate the marginal density accurately with a non-parametric \emph{f-modeling} approach\citep{efron2014two}.

Using real experiment data, the performance of Laplace prior and Huber prior differed from the simulation studies, indicating that the simulated priors and the true prior distributions are still different. Nevertheless, Ghidorah performed exceedingly well in all cases, making it a very robust and dependable choice. 

\begin{table}[!tb]
\caption{Evaluation on two metrics from real experiments. Bold numbers are the best results or very close to the best.}
\label{tab:empirical}
\resizebox{\columnwidth}{!}{%
\begin{tabular}{lllllllllll}
{\ul } &                     &                                      &       & \multicolumn{3}{c}{User Engagement Metric}                     &       & \multicolumn{3}{c}{Site Performance Metric}             \\ \cline{2-11} 
       & Method              & \multicolumn{1}{l|}{$p-$val}         & Count & RMSE          & Coverage        & \multicolumn{1}{l|}{$Var_S$} & Count & RMSE          & Coverage        & $Var_S$               \\ \cline{2-11} 
       & Unadjusted          & \multicolumn{1}{l|}{\textless{}0.01} & 34    & 5.61          & 82.4\%          & \multicolumn{1}{c|}{-}       & 40    & 7.56          & 77.5\%          & \multicolumn{1}{c}{-} \\
       &                     & \multicolumn{1}{l|}{\textless{}0.05} & 68    & 6.38          & 80.9\%          & \multicolumn{1}{c|}{-}       & 75    & 6.89          & 81.3\%          & \multicolumn{1}{c}{-} \\
       &                     & \multicolumn{1}{l|}{All}             & 519   & 4.55          & 93.4\%          & \multicolumn{1}{c|}{-}       & 522   & 4.59          & 94.3\%          & \multicolumn{1}{c}{-} \\ \cline{2-11} 
         & CMLE           & \multicolumn{1}{l|}{\textless{}0.01} & 34    & 4.54 & 100\% & \multicolumn{1}{l|}{5.36}    & 40    & {6.52} & 97.5\%          & 5.20                  \\
       &                     & \multicolumn{1}{l|}{\textless{}0.05} & 68    &  3.88& 100\% & \multicolumn{1}{l|}{5.31}    & 75    & {5.27} & {98.7\%} & 5.20                  \\
       &                     & \multicolumn{1}{l|}{All}             & 519   &  3.44&     100\%   & \multicolumn{1}{l|}{4.06}    & 522   & {3.62} & 99.8\%          & 4.09                 \\ \cline{2-11} 
       & Ghidorah            & \multicolumn{1}{l|}{\textless{}0.01} & 34    & \textbf{3.79} & \textbf{91.2\%} & \multicolumn{1}{l|}{1.00}    & 40    & \textbf{4.67} & 92.5\%          & 1.00                  \\
       &                     & \multicolumn{1}{l|}{\textless{}0.05} & 68    & \textbf{3.33} & \textbf{94.1\%} & \multicolumn{1}{l|}{0.95}    & 75    & \textbf{3.93} & \textbf{94.7\%} & 0.92                  \\
       &                     & \multicolumn{1}{l|}{All}             & 519   & \textbf{2.73} & 92.7\%          & \multicolumn{1}{l|}{0.21}    & 522   & \textbf{2.55} & 94.8\%          & 0.32                  \\ \hline
       & Normal              & \multicolumn{1}{l|}{\textless{}0.01} & 34    & 4.92          & 82.4\%          & \multicolumn{1}{l|}{0.91}    & 40    & 6.74          & 85.0\%          & 0.83                  \\
       &                     & \multicolumn{1}{l|}{\textless{}0.05} & 68    & 5.47          & 85.3\%          & \multicolumn{1}{l|}{0.89}    & 75    & 5.82          & 88.0\%          & 0.83                  \\
       &                     & \multicolumn{1}{l|}{All}             & 519   & 4.05          & 93.8\%          & \multicolumn{1}{l|}{0.86}    & 522   & 3.91          & 95.6\%          & 0.79                  \\ \hline
       & Laplace             & \multicolumn{1}{l|}{\textless{}0.01} & 34    & 4.61          & 88.2\%          & \multicolumn{1}{l|}{0.99}    & 40    & 5.85          & 90.0\%          & 0.98                  \\
       &                     & \multicolumn{1}{l|}{\textless{}0.05} & 68    & 4.97          & 91.2\%          & \multicolumn{1}{l|}{0.96}    & 75    & 5.09          & 90.7\%          & 0.94                  \\
       &                     & \multicolumn{1}{l|}{All}             & 519   & 3.63          & 94.6\%          & \multicolumn{1}{l|}{0.74}    & 522   & 3.39          & 96.0\%          & 6.94                  \\ \cline{2-11} 
       & Huber               & \multicolumn{1}{l|}{\textless{}0.01} & 34    & 4.04          & \textbf{91.2\%} & \multicolumn{1}{l|}{0.90}    & 40    & 4.91          & \textbf{95.0\%} & 0.91                  \\
       &                     & \multicolumn{1}{l|}{\textless{}0.05} & 68    & 3.63          & 92.6\%          & \multicolumn{1}{l|}{0.77}    & 75    & 4.07          & \textbf{94.7\%} & 0.80                  \\
       &                     & \multicolumn{1}{c|}{All}             & 519   & 2.91          & 94.2\%          & \multicolumn{1}{l|}{0.42}    & 522   & 2.72          & 95.4\%          & 0.45                  \\ \cline{2-11} 
       & LaplaceFitGhidorah               & \multicolumn{1}{l|}{\textless{}0.01} & 34    & 4.03          & \textbf{91.2\%} & \multicolumn{1}{l|}{0.91}    & 40    & 5.02          & \textbf{95.0\%} & 0.85                  \\
       &                     & \multicolumn{1}{l|}{\textless{}0.05} & 68    & 3.63          & 92.6\%          & \multicolumn{1}{l|}{0.78}    & 75    & 4.09          & 93.3\% & 0.72                  \\
       &                     & \multicolumn{1}{c|}{All}             & 519   & 2.90          & 94.2\%          & \multicolumn{1}{l|}{0.42}    & 522   & 2.65          & 94.4\%          & 0.36                  \\ \cline{2-11}
       & RwES(linear)        & \multicolumn{1}{c|}{\textless{}0.01} & 34    & 8.89          & 79.4\%          & \multicolumn{1}{c|}{-}       & 40    & 10.03         & 72.5\%          & \multicolumn{1}{c}{-} \\
       &                     & \multicolumn{1}{l|}{\textless{}0.05} & 68    & 7.21          & 88.2\%          & \multicolumn{1}{c|}{-}       & 75    & 7.48          & 82.7\%          & \multicolumn{1}{c}{-} \\
       &                     & \multicolumn{1}{c|}{All}             & 519   & 4.08          & 96.7\%          & \multicolumn{1}{c|}{-}       & 522   & 3.88         & 96.6\%          & \multicolumn{1}{c}{-} \\ \cline{2-11} 
       & TARwES              & \multicolumn{1}{l|}{\textless{}0.01} & 34    & 5.92          & 85.3\%          & \multicolumn{1}{c|}{-}       & 40    & 5.40          & \textbf{95.0\%} & \multicolumn{1}{c}{-} \\
       &                     & \multicolumn{1}{l|}{\textless{}0.05} & 68    & 5.17          & 91.2\%          & \multicolumn{1}{c|}{-}       & 75    & 4.34          & 96.0\%          & \multicolumn{1}{c}{-} \\
       &                     & \multicolumn{1}{l|}{All}             & 519   & 3.42          & 97.7\%          & \multicolumn{1}{c|}{-}       & 522   & 2.73          & 98.5\%          & \multicolumn{1}{c}{-} \\ \cline{2-11} 
       & TARwES+             & \multicolumn{1}{l|}{0.01}            & 34    & \textbf{3.81}          & \textbf{91.2\%} & \multicolumn{1}{c|}{-}       & 40    & 4.92          & \textbf{95.0\%} & \multicolumn{1}{c}{-} \\
       &                     & \multicolumn{1}{l|}{0.05}            & 68    & \textbf{3.35}          & \textbf{95.6\%} & \multicolumn{1}{c|}{-}       & 75    & 4.12          & 96.0\%          & \multicolumn{1}{c}{-} \\
       &                     & \multicolumn{1}{l|}{All}             & 519   & \textbf{2.73} & 97.9\%          & \multicolumn{1}{c|}{-}       & 522   & 2.60          & 98.5\%          & \multicolumn{1}{c}{-} \\ \hline
       & LocalH1(1:1)        & \multicolumn{1}{l|}{\textless{}0.01} & 34    & 5.32          & 82.4\%          & \multicolumn{1}{l|}{1.00}    & 40    & 7.27          & 85.0\%          & 1.00                  \\
       &                     & \multicolumn{1}{l|}{\textless{}0.05} & 68    & 5.50          & 86.8\%          & \multicolumn{1}{l|}{1.00}    & 75    & 6.20          & 89.3\%          & 1.00                  \\
       &                     & \multicolumn{1}{l|}{All}             & 519   & 3.50          & 95.0\%          & \multicolumn{1}{l|}{0.63}    & 522   & 3.57          & 95.6\%          & 0.66                  \\ \cline{2-11} 
       & Left:  LocalH1(1:7) & \multicolumn{1}{l|}{\textless{}0.01} & 34    & 4.34          & 88.2\%          & \multicolumn{1}{l|}{1.00}    & 40    & 6.40          & 90.0\%          & 1.00                  \\
       & Right:LocalH1(1:6)  & \multicolumn{1}{l|}{\textless{}0.05} & 68    & 3.80          & \textbf{94.1\%} & \multicolumn{1}{l|}{1.00}    & 75    & 5.19          & 90.7\%          & 1.00                  \\
       &                     & \multicolumn{1}{l|}{All}             & 519   & 2.82          & 93.3\%          & \multicolumn{1}{l|}{0.30}    & 522   & 2.93          & 94.1\%          & 0.36 \\ \hline
       & NEST        & \multicolumn{1}{l|}{\textless{}0.01} & 34    & 5.08          & 82.4\%          & \multicolumn{1}{l|}{0.97}    & 40    & 6.81          & 82.5\%          & 0.94                  \\
       &                     & \multicolumn{1}{l|}{\textless{}0.05} & 68    & 5.69          & 85.3\%          & \multicolumn{1}{l|}{0.95}    & 75    & 6.13          & 86.7\%          & 0.92                 \\
       &                     & \multicolumn{1}{l|}{All}             & 519   & 4.13          & 93.8\%          & \multicolumn{1}{l|}{0.87}    & 522   & 4.18         & 95.6\%          & 0.88                  \\ \cline{2-11} 
       
\end{tabular}%
}
\end{table}



\section{Concluding Remarks}
\label{sec:conclusion}

In the big data era, A/B testing has become an integral step in modern software development. To facilitate data-driven decision making, we conduct more and more experiments, build more and more metrics, and drill down to more and more segments. While enjoying the rich insights revealed by such practices, we should not blindly trust the results produced. In particular, we should not screen in an ad-hoc fashion, based on statistical significance, and take the filtered results by their face value. Instead, the key question that both researchers and practitioners should ask is -- given the ubiquity of post-selection in A/B testing, how do we make post-selection adjustments in a way that is both theoretically sound and practically feasible?

This paper provided an extensive discussion on post-selection inference in A/B testing, with comprehensive literature reviews, novel proposals and empirical studies. First, we raised awareness in the A/B testing community about this important and challenging problem, by highlighting that untrustworthy post-selection inference might result in biased treatment effect estimations and under-covered confidence intervals. Moreover, trustworthy post-selection inference appeared to be highly non-trivial in the context in A/B testing, due to various reasons such as heterogeneous historical experiments, the need for non-linear adjustments, the lack of ground truth for evaluation, and the possibility of limited training data. Second, after comprehensively surveying two lines of existing methods -- empirical Bayes (EB) and regression with experiment splitting (RwES), we provided two new methods, TARwES and Ghidorah. Simulation and empirical studies confirmed that Ghidorah was a robust, adaptive and training data efficient empirical Bayes method. In particular, we found that Ghidorah can be trained with as few as 50 historical experiments. Moreover, we could further improve upon Ghidorah's already outstanding performance, by combining it with TARwES. Finally, for cold start scenarios with no training data, we proposed a local Bayes Bound based method, which, with proper domain knowledge of the prior odds of the null and alternative hypotheses, could serve as a good starting point of trustworthy post-selection inference of A/B tests. We shared our implementation of all methods described in the paper, for the purpose of reproduction and adoption by the community.

There are multiple future research directions based on the current work. First, we will apply the proposed methodologies to an even broader set of real-life experiments, and search for examples where the hybrid approach TARwES+ significantly out-performs others. Second, although in this paper we focus on estimation and inference (which are more important from a business perspective), it would be interesting to deeply connect our proposals with class multiple testing adjustment procedures such as false discovery rate \cite{benjamini1995controlling, benjamini2010simultaneous}. Third, it is important to better understand the log-concave region for the Ghidorah prior from a theoretical perspective, which might lead to more robust and efficient solutions. Fourth, it is possible to generalize the current work to more complex causal inference scenarios such as non-randomized observational studies or factorial designs.

\nocite{studentt,AzevedoDeng2018}

\balance
\bibliographystyle{ACM-Reference-Format}
\bibliography{sigproc}

\clearpage

\appendix

\section{Appendix for reproducibility}

We provide R implementations of all models described and the reproduction of all simulation study results. This supplementary section details the steps to reproduce our results. Real experiments data used in empirical study are sensitive and not disclosed. 

\subsection{Reproduce simulation study}
Open the \emph{replication.ipynb} file. Run this notebook to reproduce results in Section~\ref{sec:simulation}. Make sure to run the code in order so the same random seed will apply. 

In the first cell in the notebook we install xgboost package for RwES with GBT. Installation may take a while. 

\subsection{Simulation code}
All code for simulation reside in \emph{simulation/simulation.R}. The simulation entry point is \emph{evaluateSim}. 

This function takes another function called priorGen to simulate from different ground-truth prior distributions. It also takes a list of methods to evaluate. The list of methods is simply a vector of R6 class instances for post-selection inference methods described in this paper. This function does the main evaluation loop to simulate training and testing data, run all methods to compute RMSE, variance reduction rate and interval coverages. It uses common interfaces implemented by each method to train and predict. 

Function \emph{simulateData} is the main simulation function implementing the simulation setup. 

\subsection{Code Organization}
All methods are organized in the methods folder, with their own subfolders shown in Table~\ref{tab:codeorg}. All methods implemented the same interface using R6\footnote{R6 documentation \url{https://r6.r-lib.org/}}: 
\begin{itemize}
    \item An \emph{initialize} function (constructor) which takes a training data. To initialize a R6 class, use \$new(trainingdata) syntax.
    \item A \emph{train} function to fit the model on the training data to get hyper-parameters. For methods do not train on training data, this function is no-op.
    \item A \emph{predict} function to take a test data and predict the true treatment effect. When \emph{includeVar} is set to be true, \emph{predict} also outputs the estimated variance for the prediction used for confidence interval. 
\end{itemize}

Most methods we implemented are R6 classes to be initiated by a training data. For methods that require other initialization parameters, such as oracle Bayesian posterior using known priors as the best case comparison, they are functions that return a R6 instance. For example, \emph{KnownNormalPrior(sigma)} is an instance implementing a Gaussian prior with standard deviation \emph{sigma}.

\begin{table}[h]
\caption{Methods with corresponding R6 class names and locations in source code}
\label{tab:codeorg}
\resizebox{0.5\textwidth}{!}{%
\begin{tabular}{@{}lll@{}}
\toprule
Folder                         & Methods                 & R6class                                    \\ \midrule
methods/BaselineUnadjusted     & Unadjusted MLE          & baselineUnadjust                           \\
methods/BFBound                & localH1 bound           & localH1Bound(priorOdds=)                               \\
methods/EBHuberSURE            & Huber EB                & HuberPriorEB                               \\
methods/EBmle                  & Gaussian and Laplace EB & NormalPriorEB, LaplacePriorEB            \\
methods/knownprior             & Theoretical Posterior   & Known*Prior (* are priors)                  \\
methods/MixturePrior           & Ghidorah                & ZeroNormalLaplaceMixEB                     \\
methods/poostselect            & CMLE                    & postSelectZCut(zcut=)                             \\
methods/nest            & NEST                   & NEST                             \\
\multirow{2}{*}{methods/split} & RwES                    & simpleSplitting(Linear), bstSplitting(GBT) \\
                               & TARwES                  & TARwESSplitting,GhidorahSplitting(TARwES+) \\ \bottomrule
\end{tabular}%
}
\end{table}

\end{document}